# The impact of stereotype threat on gender gap in introductory physics

Alexandru Maries[1], Nafis I. Karim[2], and Chandralekha Singh[2]

[1] *Department of Physics, University of Cincinnati, Cincinnati, OH 45221*
[2] *Department of Physics and Astronomy, University of Pittsburgh, Pittsburgh, PA 15260*

Many prior studies have found a gender gap between male and female students' performance on conceptual assessments such as the Force Concept Inventory (FCI) and the Conceptual Survey of Electricity and Magnetism (CSEM) with male students performing better than female students. Prior studies have also found that activation of a negative stereotype about a group or stereotype threat, e.g., asking test-takers to indicate their ethnicity before taking a test, can lead to deteriorated performance of the stereotyped group. Here, we describe two studies in which we investigated the gender gap on the FCI and CSEM. In the first study, we investigated whether asking students to indicate their gender immediately before taking the CSEM increased the gender gap compared to students who were not asked for this information. In the second study, conducted with over 1100 introductory physics students, we investigated the prevalence of the belief that men generally perform better in physics than women and the extent to which this belief is correlated with the performance of both the female and male students on the FCI.

## I. INTRODUCTION

In introductory physics, prior research has found that male students often outperform female students on conceptual assessments such as the Force Concept Inventory or FCI [1,2] and the Conceptual Survey of Electricity and Magnetism or CSEM [3], a phenomenon sometimes referred to as the "gender gap". Furthermore, prior research has also found that activation of a stereotype about a particular group in a test-taking situation, i.e., stereotype threat, can alter the performance of that group in a way consistent with the stereotype. For example, Spence et al. [4] conducted a study in which a group of students was told before taking a math test that in prior administrations of the test, a gender gap has been found (with female students performing worse than male students), while another group was not provided with this information. Female students who were told the stereotypical statement performed significantly worse than those who were not exposed to this stereotype threat (with such a statement), but the performance of male students was unaffected. Other researchers have found more subtle stimuli that can activate stereotype threat and result in deteriorated performance [5], e.g., asking students to indicate their ethnicity before taking a test [6]. In particular, prior research suggests that asking African American students to indicate their ethnicity before taking a difficult test on verbal ability resulted in decreased performance compared to students who were not [6]. Yet others have found that asking for gender or ethnicity before taking a test does not impact students' performance on standardized tests [7,8].

Since stereotype threat has the potential to exacerbate the gender gap typically found in conceptual physics assessments, in Study 1, we investigated whether asking introductory students to indicate their gender before taking the CSEM impacted their performance, both when it was administered as a pre-test (before instruction) and as a post-test (after instruction in relevant concepts). In Study 2, we investigated the prevalence of the belief that men generally perform better in physics than women (a gender stereotype) among introductory students and the extent to which agreeing with this gender stereotype is correlated with the performance of female and male students on the FCI. We hypothesized that asking students for their beliefs about this gender stereotype may act as a stereotype threat, especially for female students who agree with the stereotype, and they may perform worse than female students who do not agree with it. The studies were conducted in two consecutive years.

## II. METHODOLOGY

*Study 1*: In this study, 170 students in an introductory algebra-based physics II course (mostly pre-medical students and bioscience majors) took the CSEM as a pre-test (in the first week of classes) and as a post-test (last week of classes). Students were randomly assigned to two conditions, one which asked them to indicate their gender (checkbox format with options male, female, and prefer not to specify) and one which did not immediately before taking the CSEM. We then compared the performance of students in the two conditions.

*Study 2*: In this study, we investigated the following: 1) the prevalence of the belief in the gender stereotype among introductory physics students and 2) the extent to which believing the stereotype is correlated with female and male students' performance on the FCI. This study involved over 1100 students enrolled in first semester algebra- and calculus-based (mainly engineering and physical science majors) introductory physics courses. Students were asked to indicate the extent to which they agree with the following statement: "According to my own personal beliefs, I expect men to generally perform better in physics than women" on a five point Likert scale (strongly disagree, disagree, neutral, agree, and strongly agree). Then, we grouped students according to their beliefs (agree/strongly agree, neutral, disagree/strongly disagree) and looked for performance differences (e.g., comparing the performance of female students who agree with the stereotype with that of female students who disagree with the stereotype) on both the pre-test and the post-test.

## III. RESULTS

Study 1: Figure 1 shows the pre-test and post-test performance of introductory algebra-based female (N=99) and male (N=71) students on the CSEM in the two conditions: students were/were not asked to provide gender information before taking the CSEM (gender salient/not salient condition). Figure 1 shows that there were no statistically significant differences between male or female students in the two conditions for the pre-test and post-test.

Study 2: Table I shows the percentage of male and female introductory students in algebra-based and calculus-based physics courses who agreed/were neutral/disagreed with the stereotype (I expect men to generally perform better in physics than women). Only approximately 8% of algebra-based students and 10% of calculus-based students agreed with the stereotype. Thus, it appears that the stereotype was not very common amongst introductory physics students.

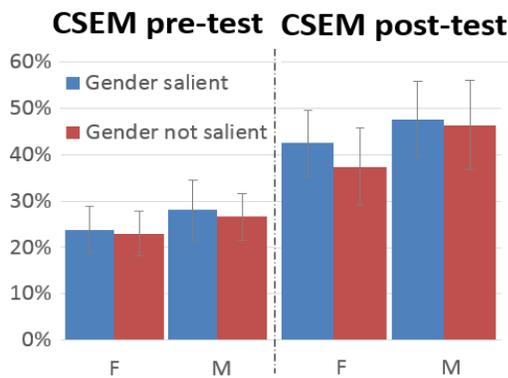

FIG. 1. Female (F) and male (M) students' pre- and post-test performance on the CSEM based on testing condition. The bars indicate standard deviations.

TABLE I. Percentage of female (F) and male (M) students who agreed/were neutral/disagreed with the stereotype that men generally perform better in physics than women in algebra-based (Alg.) and calculus-based (Calc.) introductory physics. The numbers (N) in parentheses indicate the total number of female/male students.

|      |           | Agree | Neutral | Disagree |
|------|-----------|-------|---------|----------|
| Alg. | F (N=210) | 8.6%  | 16.2%   | 75.2%    |
|      | M (N=107) | 7.5%  | 24.3%   | 68.2%    |
| Calc.| F (N=177) | 11.3% | 7.3%    | 81.4%    |
|      | M (N=651) | 9.5%  | 33.9%   | 56.5%    |

Table II shows the pre-test and post-test performances on the FCI of female and male students in algebra- and calculus-based physics who agreed/disagreed with the stereotype. Table II also lists *p* values and effect sizes (Cohen's d) for the following comparisons (the *p* values listed were obtained via *t*-tests and are not listed whenever the number of students in one group was less than 14 because *t*-tests are not appropriate for such small numbers):

- *Female students who agreed or disagreed with the stereotype compared to male students who agreed or disagreed with the stereotype*: These are listed on the far right under each class type. For example, in the pre-test for calculus-based physics, female students who agreed with the stereotype performed significantly worse on the FCI than male students who agreed with the stereotype ($p$ value < 0.001, effect size = 1.36).
- *Female/male students who agreed with the stereotype compared to students of the same gender who disagreed with the stereotype*: These are shown underneath the performance of female/male students for each class type. For example, in the pre-test for calculus-based physics, female students who agreed with the stereotype performed significantly worse than female students who disagreed with the stereotype ($p$ value = 0.003, effect size = 0.69).
- The effect sizes are always listed as positive. Which group performed better can easily be determined from the average performance of the groups being compared.

Table II shows that on the pre-test for both calculus- and algebra-based students, the average performance of female students who agreed with the stereotype was significantly lower than the average performance of female students who did not agree with the stereotype. The difference in performance between these two groups of female students appears to be larger in the calculus-based class than in the algebra-based class (10% in calculus-based compared to only 4% in the algebra-based). On the post-test for calculus-based introductory physics, it appears that female students who agreed with the stereotype continued to perform worse than female students who disagreed, although the numbers are too small to perform a *t*-test. However, the effect size (0.39) suggests that if the number of female students was larger, the difference may have possibly become statistically significant. In the post-test for algebra-based introductory physics, there was no difference in performance between female students who agreed and female students who disagreed with the stereotype. Thus, it appears stereotype threat may have a more detrimental effect for female students in calculus-based physics compared to female students in algebra-based physics. Possible reasons for this finding are discussed in the discussion and summary section.

**Table II.** Numbers of algebra-based and calculus-based students (N), averages (Av) and standard deviations (Sd) for the performance on the FCI (pre-/post-test) of female and male students who agree (A) and disagree (D) with the stereotype that men generally perform better in physics than women. The *p* values (*p* val.) and effect sizes (E.s.) in the middle of the table and on the right were obtained when comparing the average performance of female students who agree/disagree with that of male students who agree/disagree with the stereotype. The *p* values and effect sizes shown underneath the performance of female/male students for each class type were obtained when comparing the average performance of female/male students who agree with that of female/male students who disagree with the stereotype. For any group comparisons with fewer than 14 students, *p* values are not listed due to the small numbers.

| | Calculus based pre-test | | | | | | | | Algebra based pre-test | | | | | | | |
|---|---|---|---|---|---|---|---|---|---|---|---|---|---|---|---|---|
| | Female students | | | Male students | | | | | Female students | | | Male students | | | | |
| | N | Av | Sd | N | Av | Sd | *p* val. | E.s. | N | Av | Sd | N | Av | Sd | *p* val. | E.s |
| A | 20 | 30 | 13 | 62 | 53 | 20 | <0.001 | 1.36 | 18 | 28 | 8 | 8 | 52 | 18 | | 1.78 |
| D | 145 | 40 | 18 | 368 | 54 | 20 | <0.001 | 0.71 | 158 | 32 | 17 | 72 | 43 | 19 | <0.001 | 0.62 |
| | *p* val. = 0.003 | | | *p* val. = 0.792 | | | | | *p* val. = 0.049 | | | | | | | |
| | E.s. = 0.69 | | | E.s. = 0.04 | | | | | E.s. = 0.36 | | | E.s. = 0.47 | | | | |
| | Calculus based post-test | | | | | | | | Algebra based post-test | | | | | | | |
| | Female students | | | Male students | | | | | Female students | | | Male students | | | | |
| | N | Av | Sd | N | Av | Sd | *p* val. | E.s. | N | Av | Sd | N | Av | Sd | *p* val. | E.s |
| A | 10 | 49 | 17 | 44 | 68 | 20 | 0.007 | 1.02 | 14 | 46 | 17 | 6 | 47 | 18 | | 0.27 |
| D | 117 | 55 | 18 | 273 | 68 | 20 | <0.001 | 0.65 | 136 | 47 | 18 | 60 | 59 | 20 | <0.001 | 0.66 |
| | | | | *p* val. = 0.980 | | | | | *p* val. = 0.910 | | | | | | | |
| | E.s. = 0.39 | | | E.s. = 0.00 | | | | | E.s. = 0.03 | | | E.s. = 0.65 | | | | |

For male students in calculus-based physics, there appear to be no differences between male students who disagree with the stereotype and male students who agree, whereas in algebra-based physics, the numbers are too small to draw any conclusions.

Furthermore, Table II shows that the largest differences in performance between male and female students were between male and female students who agreed with the stereotype. Three out of the four comparisons (calculus-based pre-test and post-test, algebra-based pre-test) resulted in large effect sizes (1.36, 1.02, and 1.78, respectively).

## IV. DISCUSSION AND SUMMARY

In Study 1, our research suggests that asking algebra-based introductory physics students to indicate their gender before taking the CSEM did not impact their performance, consistent with a previous study conducted with the AP calculus exam and the Computerized Placement test [7]. One possible explanation for this finding supported by previous research [9] is that stereotype threat for female students occurs regardless of whether or not students are asked to indicate their gender before taking the CSEM test because the stereotype is automatically activated for female students in the test-taking situation. Other high-stakes tests (e.g., MCAT, SAT) commonly require students to indicate their gender before taking the tests. If the results of Study 1 were to hold for these tests as well, then the common practice of asking for personal information such as gender may not make a difference in the performance of the stereotypically underperforming group. One possible explanation is that the threat may be present for this group regardless of being asked about such personal information explicitly.

In Study 2, we investigated the prevalence of the belief that men generally perform better in physics than women among introductory physics students and found that this type of belief is not very common (about 8% of algebra-based students and 10% of calculus-based students agree with this stereotype). Moreover, we also investigated the extent to which agreeing with the stereotype was correlated with students' performance on the FCI and found that in some situations, female students who agreed with the stereotype performed worse than female students who did not agree with it. This effect appears to be stronger in the calculus-based courses compared to the algebra-based courses. It is unclear why this is the case, but one possible reason which may partly account for this difference is the lower percentage of female students in introductory calculus-based physics courses compared to algebra-based courses. In our data, female students comprised 63% of the students in the algebra-based courses (primarily pre-medical and biological science majors), but only 21% of the students in the calculus-based courses (primarily engineering, mathematics and chemistry majors). Thus, in a calculus-based course, female students who agree with the stereotype are likely to be impacted more by the stereotype threat since they see fewer

258

female students compared to male students in their physics class. In other words, the observation that there are fewer female students in a physics class compared to male students can reinforce the stereotype and hence has the potential to cause a larger stereotype threat. This could lead to increased anxiety for female students in a test-taking situation in a calculus-based course compared to an algebra-based course. In an algebra-based course, the observation of larger percentage of female students may be perceived by students as inconsistent with the stereotype and therefore may potentially reduce the impact of the stereotype threat even for the female students who agree with the stereotype. Finally, we found that the gender gap is the largest between male and female students who agree with the stereotype, often with very large effect sizes.

Finally, we should note that the results of this investigation can be useful for designing professional development for instructors and TAs to help them make their classes more inclusive [10,11]. Our data indicate that agreeing with the gender stereotype that men generally perform better in physics than women is correlated with decreased performance for female students. One possible explanation of this finding is that female students who agree with the stereotype may experience increased stereotype threat than female students who do not agree with the stereotype. Thus, TAs and instructors need to be careful to not propagate these types of stereotypes, both in their actions and statements. In particular, instructors and TAs should try to send the message to their students (both explicitly and implicitly) that success in physics is primarily determined by effort and engaging in appropriate learning strategies rather than by something innate, e.g., gender (i.e., they should send the message that all students regardless of their gender can excel by effort and deliberate practice). In a book chapter titled "Is Math a Gift? Beliefs that Put Females at Risk" [12], Dweck argues that a fixed mindset (belief that intelligence is fixed or innate) is more detrimental to female students than male students. She describes a study in which two groups of adolescents were taught the same math lesson (which included historical information about the mathematicians who originated the ideas discussed in the lesson) in two different ways. For one group, the mathematicians were portrayed as geniuses and their "innate ability" and "natural talent" were highlighted, whereas for the other group, the mathematicians' commitment and hard work were highlighted. After the lesson, students were given a difficult math test and were told that the test would measure their mathematical ability. Female students who received the lesson which portrayed the mathematicians as geniuses performed worse than their male counterparts. On the other hand, for students who received the lesson which highlighted the mathematicians' hard work, there were no gender differences in performance. Dweck argues that when female students receive messages that mathematics ability is a gift, some of them may interpret that this gift is something they do not possess [12]. It is possible that accumulated societal stereotypes influence how female students interpret these messages and they may assume that if mathematical ability is a gift, male students are likely to have this gift, whereas they are not likely to have it. Therefore, it is important that professional development workshops for physics instructors and TAs focus on the findings of this research vis a vis other studies on stereotype threat [9,10,12], and help instructors and TAs reflect upon the importance of encouraging their students to develop a growth mindset, namely that intelligence is malleable and it can be cultivated with hard work and productive learning strategies regardless of gender or other characteristics (e.g., race/ethnicity) of an individual.